\documentclass{elsart}

\usepackage{graphicx}
\usepackage{amssymb}

\begin{document}

\begin{frontmatter}
\date{}
\title{Influence of the vertical closed orbit distortions on accuracy of the energy calibration done by resonant depolarization technique\thanksref{ref}.}
\thanks[ref]{The work was supported in part by grant of Russian Fund for Basic Research No. 01-02-17477.}

\author{A.V.~Bogomyagkov\thanksref{cor1},}
\thanks[cor1]{corresponding author, e-mail:  A.V.Bogomyagkov@inp.nsk.su    }
\author{S.A.~Nikitin,}
\author{A.G.~Shamov.}
\address{The Budker Institute of Nuclear Physics,
    Acad. Lavrentiev prospect 11,\\
    630090 Novosibirsk,
    Russia }

\begin{abstract}
The series of the experiments on precise mass measurement of $J/\Psi$-, $\Psi'$- mesons have been performed in 2002-2004. Energy calibration has been done with the help of the resonant depolarization technique. The present paper discusses the influence of the vertical orbit distortions on the accuracy of the energy calibration. The sources of the orbit distortions are misalignments of the quadrupoles and sextupoles in vertical plane and kicks of the vertical correctors. Comparison with previously published papers is presented.

\end{abstract}
\end{frontmatter}

\section{Introduction}
The series of the experiments on precise mass measurement of $J/\Psi$-, $\Psi'$- mesons have been performed in 2002-2004. The following mass values have been obtained \cite{meson-masses}:
\[
M_{J/\Psi} = 3096.917\pm0.010\pm0.007 \;MeV\,,
\]
\[
M_{\Psi^\prime}= 3686.111\pm0.025\pm0.009 \;MeV\,.
\]
Energy calibration of the colliding beams has been performed by resonant depolarization technique. To achieve high accuracy of the mass measurements an analysis of possible errors have been performed \cite{PAC2001,analysis-of-errors}. In particular, the effect of the vertical closed orbit distortions influence on accuracy of the energy calibration was preliminary estimated with the result of $14$~keV correction for the mass of $\Psi'$--meson. Comparatively large value of possible energy bias stimulated the further analysis which is presented below.

The discussed effect have been addressed by numerous authors \cite{MacKay,Baru,Shatunov,Assmann}. The common conclusion was that value of the systematic error in energy calibration is proportional to the squared value of the orbital distortions. Initially we have performed similar calculations and compared them with simulation. The difference by order of magnitude was observed at integer resonance vicinity and also the theoretical estimation predicted the opposite sign of the effect value at the energy region of $\Psi'$-meson.

The authors of \cite{Shatunov} proposed usage of single spin harmonic (measured, for example, by polarization life time) for evaluation the energy bias. This approach was compared with simulation and limitations for usage were found.

The present paper discusses all mentioned approaches and gives more accurate estimation of the effect.

\section{The problem definition}
The spin precession frequency $\Omega_S$ of the particle moving in vertical guiding field is described by
\begin{equation}
\Omega_S=\Omega_0(1+\gamma\frac{q^\prime}{q_0}) \,,
\label{eq:OmegaS}
\end{equation}
where $\Omega_0=q_0B/\gamma$ is revolution frequency, $B$ is average guiding field, $\gamma$ is Lorentz factor, $q^\prime$, $q_0$ are anomalous and normal parts of gyromagnetic ratio. Introducing the spin tune $\nu=(\Omega_S-\Omega_0)/\Omega_0=\gamma q^\prime/q_0$ one will have a known relation between energy $E$ and spin tune
\begin{equation}
E[MeV]=\nu\times440.64843(3)\,.
\label{eq:Enu}
\end{equation}
The closed orbit in this case is assumed to be flat. In general, the radial and the longitudinal magnetic as well as vertical electric fields may exist in real accelerator that makes the given equation inadequate. In the first order of perturbation theory the modified relation between spin frequency and beam energy can be expressed in the form
\begin{equation}
\nu^\prime=\gamma\frac{q^\prime}{q_0}+\Delta\nu(\gamma,perturbations)\,.
\label{eq:nu-dnu}
\end{equation}
The goal is to estimate spin tune shift $\Delta \nu$ by the given perturbations. This gives a possibility to find the correct energy value by the quantity $\nu^\prime-\Delta \nu$, where $\nu^\prime$ is a spin tune measured by the resonant depolarization technique and the following relation has to be used:
\[
\gamma=\frac{q_0}{q^\prime}(\nu^\prime-\Delta\nu)\,.
\]

The longitudinal fields arise from the errors of compensation of the detector's field. Consideration of these perturbations is most simple and presented in \cite{PAC2001}.We consider the influence of the radial fields which arise primary due to misalignment of quadrupoles in vertical plane.

\section{The general approach}
\begin{figure}[htb]
\centering
\includegraphics*[width=150mm]{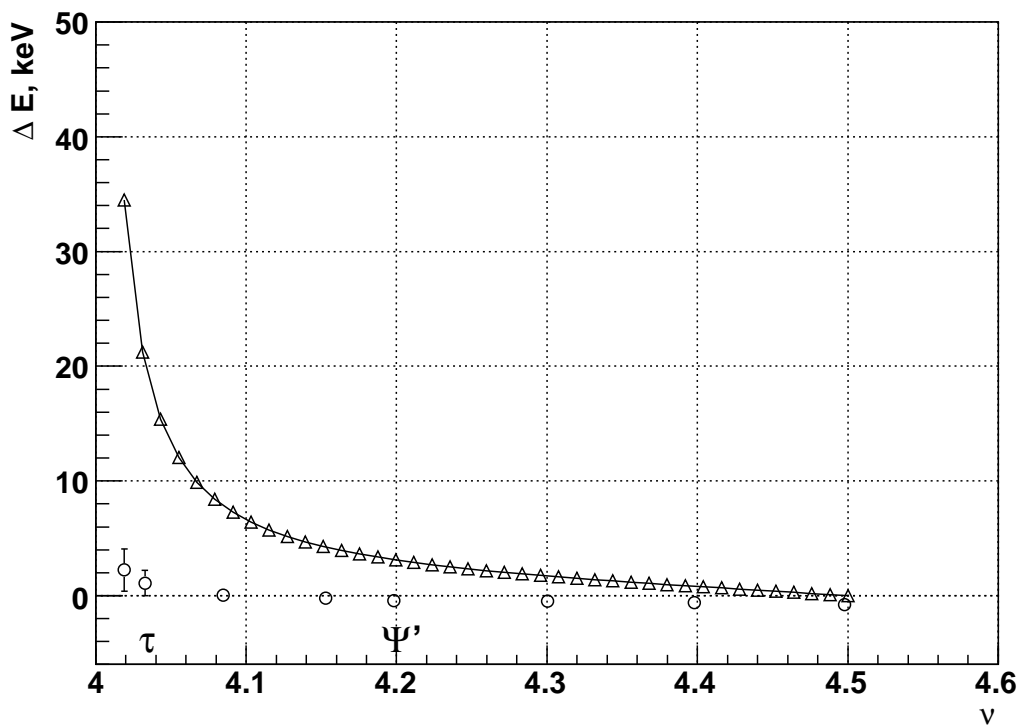}
\caption{Energy shift versus spin tune at 1~mm vertical orbit RMS. Triangles are calculations by formula (\ref{eq:Dnu-2}), circles with errors are results of the simulation.}
\label{fig:result0-0}
\end{figure}

To calculate the shift of the spin tune in the presence of the radial field we will assume sources of perturbations (including vertical correctors, quadrupole lenses and other sources of radial fields) to be point-like and rather weak. The calculations will be done in the second order of perturbation theory with the help of spinor matrices technique \cite{Derbenev} using Pauli matrices $(\sigma_x,\, \sigma_y,\, \sigma_z)$ and the unit $2\times2$ matrix $I$. Also, the coordinate system is related to the velocity vector of the equilibrium particle. Thus, rotation angle of the spin vector $2\chi=\nu\alpha$ is proportional to the rotation angle of the velocity vector $\alpha$.  The spinor matrix for the rotation around radial ($x$) basis vector on the angle $2\chi_i=\nu\alpha_i$ for the perturbation at azimuth $\theta_i$ is 
\[
T_i=I\cos(\chi_i)-i\sigma_x\sin(\chi_i)\,.
\]
Spin evolution in vertical field is described by
\[
M_i=I\cos\left(\frac{\Phi_{i+1,i}}{2}\right)-
i\sigma_z\sin\left(\frac{\Phi_{i+1,i}}{2}\right)\,,
\]
where $\Phi_{i+1,i}=\Phi(\theta_{i+1})-\Phi(\theta_i)$ and $\Phi(\theta_i)=\int\limits_0^{\theta_i}\nu Kd\theta$ is a rotation angle of the spin vector in the guiding field from the origin azimuth to given perturbation location; $K$ is the orbit curvature in units of the inverse mean machine radius $1/R$.
The total one-turn matrix of the spin evolution is obtained by multiplication of the subsequent spinor matrices
\[
M=\prod_iT_iM_i\,.
\]
The new spin tune $\nu^\prime$ is obtained from the following formula $\cos(\pi\nu^\prime)=1/2\,\mathrm{Sp}(M)$, while $\nu$ denotes spin tune without radial fields. It is simple to calculate spin tune in the case of one perturbation (neglecting higher than second order terms)
\[
\cos(\pi\nu)-\cos(\pi\nu^\prime)=\displaystyle\frac{\chi_1^2}{2}\cos(\pi\nu)\,,
\]
of two perturbations
\[
\cos(\pi\nu)-\cos(\pi\nu^\prime)=\displaystyle\frac{\chi_1^2+\chi_2^2}{2}\cos(\pi\nu)+\chi_1\chi_2\cos(\pi\nu-\Phi_{2,1})\,,
\]
of $N$ perturbations
\begin{equation}
\cos(\pi\nu)-\cos(\pi\nu^\prime)=\cos(\pi\nu)\sum_{i=1}^{N}
\displaystyle
\frac{\chi_i^2}{2}+\sum_{j>i, i=1}^{N}\chi_i\chi_j\cos(\pi\nu-\Phi_{j,i})\,.
\end{equation}
This gives the spin tune shift
\begin{equation}
\Delta\nu=\nu^\prime-\nu=\frac{1}{2\pi\sin\pi\nu}\left[
\cos\pi\nu\sum\chi_i^2+2\sum_{j>i}\chi_i\chi_j\cos(\pi\nu-\Phi_{j,i})
\right]\!.
\label{eq:Dnu}
\end{equation}
The first term in the right part of the equation describes non-correlated part of the orbital distortions influence, the second one corresponds to their correlations. The authors of \cite{MacKay,Baru,Shatunov,Assmann} neglected the second term by statistical considerations or by assumption that closed orbit is well corrected and remained distortions are statistically independent. Hence, the spin tune shift is
\begin{equation}
\Delta\nu=\frac{\cos\pi\nu}{2\pi\sin\pi\nu}\sum\chi_i^2=
          \frac{\nu^2\cot\pi\nu}{8\pi}\sum\alpha_i^2\,.
\label{eq:Dnu-1}
\end{equation}
The summation over the orbit rotation angles $\alpha_i$ was estimated by using observed vertical orbit RMS $\left<z^2\right>$ (assuming that $\left<z\right>=0$), number of quadrupole lenses $N$ (inclusion of vertical orbit correctors does not change the result much for VEPP-4M) and average focus distance of the lenses $F$. The final estimation is following
\begin{equation}
\Delta\nu=\frac{\nu^2\cot(\pi\nu)}{8\pi}\frac{N\left<z^2\right>}{F^2}\,.
\label{eq:Dnu-2}
\end{equation}
The comparison of the calculations by obtained formula with simulation is presented on Fig.\ref{fig:result0-0}, where $\Delta E=440.65\cdot\Delta\nu$.
As it could be seen the obtained estimation gives energy bias about 10 times bigger than simulation for energy region of $\tau$ lepton, sign of the estimation is opposite to one from simulation at $\Psi'$ region and value of estimation is zero at half integer spin tune when simulation value is not.

Equation (\ref{eq:Dnu}) could be written in the integral form which will be used for further calculations. Given $z$ is the vertical closed orbit deviations in units of $R$ and $z^{\prime\prime}=d^2z/d\theta^2$. The spin rotation angle is than $2\chi_i=\nu z^{\prime\prime}\Delta\theta_i$, where $\Delta\theta_i$ is an interval of the $i$-th perturbation. Thus, the spin tune shift is
%Given $z$ is the vertical closed orbit deviations in units of $R$ and $z^{\prime\prime}=d^2z/d\theta^2$. The spin rotation angle is than $2\chi_i=\nu z^{\prime\prime}\Delta\theta_i$, where $\Delta\theta_i$ is an interval of the $i$-th perturbation. Taking into account the last definitions, the equation (\ref{eq:Dnu}) could be written in the integral form:
\begin{equation}
\Delta \nu=\frac{1}{16\pi\sin{\pi \nu}}
\int\limits_0^{2\pi}\nu z^{\prime\prime}d\theta\int\limits_0^{2\pi}
\nu z^{\prime\prime}
\left[e^{i(\pi\nu-|\Phi-\Phi'|)}+c.c.\right]d\theta^\prime\,.
\label{eq:Dnu-integral}
\end{equation}
Introducing the definition of the spin harmonic amplitude
\begin{equation}
\omega_k=\frac{1}{2\pi}
\int\limits_0^{2\pi}\nu z^{\prime\prime}
\exp{[-i(\Phi-\nu\theta)-ik\theta]}d\theta\,,
\end{equation}
it is possible to transform (\ref{eq:Dnu-integral}) to the form obtained by A.M.Kon\-dra\-ten\-ko \cite{Kondratenko} (who derived this equation in different way)
\begin{equation}
\Delta\nu=\frac{1}{2}\sum\limits_k \frac{|\omega_k|^2}{\nu-k}\,.
\label{eq:kondratenko}
\end{equation}

The goal of the following calculations is to estimate spin harmonic amplitude $\omega_k$ by vertical orbit RMS $\left<z^2\right>$.

\section{Calculation of the spin harmonics}
\subsection{No straight sections and constant beta}
Assuming accelerator without straight sections i.e. $\Phi=\nu\theta$ and given Fourier expansion of $z=\sum z_n e^{in\theta}$ and $z^{\prime\prime}=-\sum z_n n^2 e^{in\theta}$ one can obtain that $\omega_k=-\nu k^2 z_k$ and corresponding spin tune shift is
\begin{equation}
\Delta\nu=\displaystyle\frac{1}{2}\sum_{k=-\infty}^{\infty}\frac{|\omega_k|^2}{\nu-k}=
\frac{\nu^2}{2}\sum_{k=-\infty}^{\infty} \frac{|z_k|^2k^4}{\nu-k}\,.
\label{eq:intermediate}
\end{equation}

To evaluate orbit harmonics $z_n$ it is convenient to use the known variables $u=z/\sqrt{\beta_z}$ and
$\phi=\int_0^\theta d\theta/(\nu_z\beta_z)$, where $\beta_z$ is vertical beta function in units of $R$, $\nu_z$ is vertical betatron tune. Then the closed orbit equation is written as:
\begin{equation}
\frac{d^2u}{d\phi^2}+\nu_z^2u=\nu_z^2\beta_z^{3/2}h(\phi)=F(\phi)\,,
\label{eq:motion}
\end{equation}
where $h(\phi)=\Delta H_x/\left<H_z\right>$.
Performing Fourier decomposition on both parts of equation (\ref{eq:motion}) one obtains
\[
u_n=\frac{F_n}{\nu_z^2-n^2}\,,
\]
where $u=\sum_{n=-\infty}^{\infty} u_n e^{i n\phi}$, $F=\sum_{n=-\infty}^{\infty} F_n e^{i n\phi}$. The RMS of orbit distortions is calculated by summation over squared harmonic amplitudes
\begin{equation}
\left<u^2\right>=\sum_{n=-\infty}^\infty |u_n|^2=
\sum_{n=-\infty}^\infty\frac{F_nF_n^*}{(\nu_z^2-n^2)^2}\,.
\end{equation}
Assuming that all orbits with the same RMS are produced by random and uniform kicks $F(\phi)$ i.e. $\overline{F_iF_j^*}=f^2\delta_{ij}$ (where $\bar{\,}$ denotes averaging over orbits with the same RMS and $^*$ is complex conjugation) we calculate the RMS orbit distortion
\begin{equation}
\overline{\left<u^2\right>}=\sum_{n=-\infty}^\infty \overline{|u_n|^2}=
f^2\sum_{n=-\infty}^\infty \frac{1}{(\nu_z^2-n^2)^2}=f^2Q\,,
\label{eq:RMS_U}
\end{equation}
where 
\begin{equation}
Q=\displaystyle\frac{\pi}{2\nu_z^3}\cot\pi\nu_z+\frac{\pi^2}{2\nu_Z^2}\csc^2\pi\nu_z\,.
\end{equation}
The obtained relation (\ref{eq:RMS_U}) allows to find mean squared excitation $f^2=\overline{\left<u^2\right>}/Q$.

In homogeneous approximation $\beta_z=const=\left<\beta_z\right>$ the following relations could be written $\phi(\theta)=\theta$, $z_n=u_n\sqrt{\left<\beta_z\right>}$, $\left<z^2\right>=\left<u^2\right>\left<\beta_z\right>$ and
\begin{eqnarray}
\overline{|z_n|^2}&=&\overline{|u_n|^2}\left<\beta_z\right>=
\left<\beta_z\right>\frac{\overline{|F_n|^2}}{(\nu_z^2-n^2)^2}= \nonumber \\
&=&\left<\beta_z\right>\frac{\overline{\left<u^2\right>}}{Q}\frac{1}{(\nu_z^2-n^2)^2}=
\frac{\overline{\left<z^2\right>}}{Q}\frac{1}{(\nu_z^2-n^2)^2}\,.
\end{eqnarray}
Substituting obtained relations into formula (\ref{eq:intermediate}) we obtain the desired relation between average spin tune shift and orbit RMS
\begin{equation}
\overline{\Delta\nu}=\displaystyle
\frac{\nu^2}{2}\frac{\overline{\left<z^2\right>}}{Q}
\sum_{k=-\infty}^{\infty} \frac{k^4}{(\nu_z^2-k^2)^2(\nu-k)}\,.
\label{eq:result0-1}
\end{equation}

\begin{figure}[htb]
\centering
\includegraphics*[width=150mm]{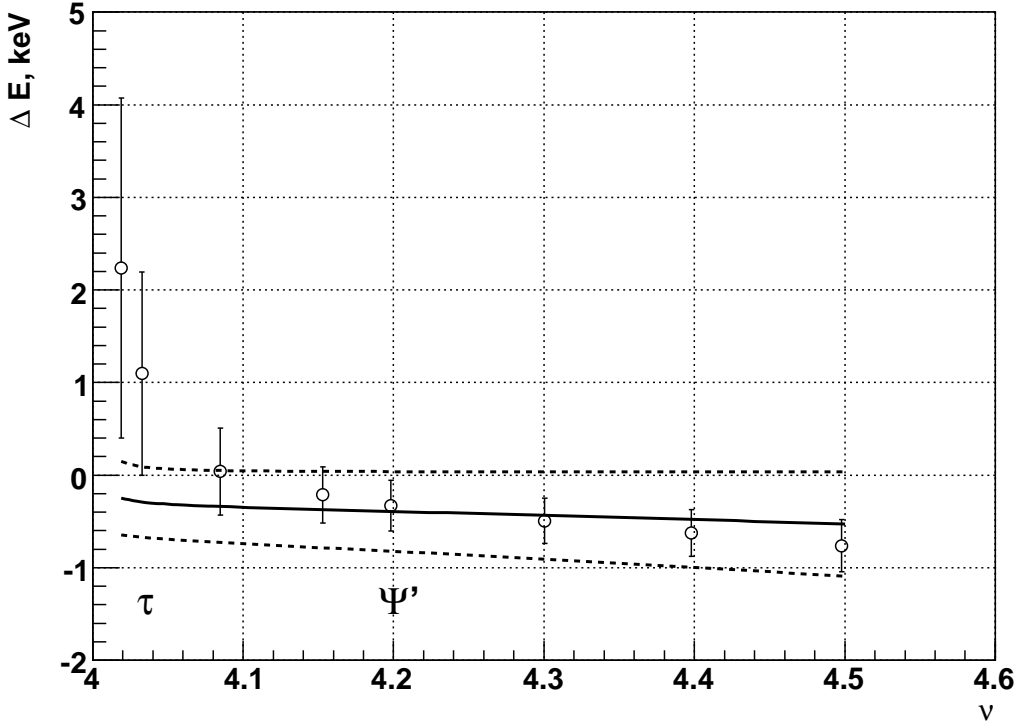}
\caption{Energy shift versus spin tune at 1~mm vertical orbit RMS. The solid line represents estimation by (\ref{eq:result0-1}), dashed lines represent the uncertainty of the estimate, calculated by (\ref{eq:result0-1-sigma}), circles with errors are results of the simulation.}
\label{fig:result0-1}
\end{figure}

To evaluate the uncertainty of the above estimation it is necessary to calculate
\begin{eqnarray}
\overline{\Delta\nu^2}&=&\left(\frac{\nu^2\left<\beta_z\right>}{2}\right)^2
\sum_{k,n=-\infty}^{\infty}
\frac{k^4}{(\nu_Z^2-k^2)^2(\nu-k)}\times \nonumber \\
& & \times\frac{n^4}{(\nu_z^2-n^2)^2(\nu-n)}
\overline{|F_k|^2|F_n|^2}\,,
\end{eqnarray}
where averaging is performed over all possible orbits with the same RMS.
Taking into account that $\overline{F_kF_k^*F_nF_n^*}=3f^4(\delta_{k,n}+\delta_{k,-n})+f^4$, we obtain
\begin{eqnarray}
\sigma_{\overline{\Delta\nu}}=\sqrt{\overline{\Delta\nu^2}-\overline{\Delta\nu}^2}&=&
\displaystyle\frac{\nu^2\sqrt{3}}{2}\frac{\left<z^2\right>}{Q}\times \nonumber \\
& &\times\sqrt{2\nu\sum_{k=-\infty}^{\infty} \frac{k^8}{(\nu_z^2-k^2)^4(\nu-k)^2(\nu+k)}}\,.
\label{eq:result0-1-sigma}
\end{eqnarray}
Comparison of the obtained estimation with simulation for VEPP-4M is presented on Fig.\ref{fig:result0-1}
Performed calculations are in a good agreement with simulation in the region far from the integer spin resonance $\nu=4$. The discrepancy between simulation and calculation in the vicinity of integer spin resonance is due to underestimated resonant spin harmonic for VEPP-4M. To perform better estimation our assumptions about absence of the straight sections and constant beta have to be changed. 
There are two long straight sections of 40~m each with circumference of 366~m at VEPP-4M.

\subsection{Straight sections and beta function variations}
The straight sections give more complicated relation between spin and orbital harmonics
\begin{eqnarray}
\omega_k&=&\frac{1}{2\pi}\int\limits_0^{2\pi}\nu_0 z^{\prime\prime}
\exp{[-i(\Phi-\nu\theta)-ik\theta]}d\theta= \nonumber \\
&=&-\frac{\nu}{2\pi}\sum_{n=-\infty}^{\infty} z_nn^2\int\limits_0^{2\pi}
\exp{[-i(\Phi-\nu\theta)-i(k-n)\theta]}d\theta= \nonumber \\
&=&-\nu\sum_{n=-\infty}^{\infty} D_{kn}z_nn^2\,,
\label{eq:harmonic-full}
\end{eqnarray}
where elements of the matrix $D_{kn}$ are calculated as
\begin{equation}
D_{kn}=\frac{1}{2\pi}\int_0^{2\pi}
\exp{[-i(\Phi-\nu\theta)-i(k-n)\theta]}d\theta\,.
\end{equation}
There are two straight sections with the length of $L$ each, in case of VEPP-4M. The straight sections are separated by arcs with radius $R$. The calculated elements of the matrix $D_{kn}$ in described layout are
\begin{equation}
D_{kn}=\frac{\nu r}{\pi\delta\Delta}\sin{\left(\frac{x_1\delta}{2}\right)}
\left[1+\cos\bigl(\pi(k-n)\bigr)\right]\,,
\end{equation}
where $\delta=k-n-\nu$, $\Delta=\nu(r-1)+k-n$, $r=R_0/R$, $x_1=L/R_0=\pi(r-1)/r$.

Azimuthal beta function variations leads to the following relations between orbital harmonic $z_n$ (along azimuth $\theta$, $z=\sum_{n=-\infty}^{\infty} z_n e^{in\theta}$) and $u_k$ (along azimuth $\varphi$, $u=\sum_{k=-\infty}^{\infty} u_k e^{i k\phi}$)
\begin{eqnarray}
z_n&=&\frac{1}{2\pi}\int_0^{2\pi}z(\theta)e^{-in\theta}d\theta=
\frac{1}{2\pi}\int_0^{2\pi}u(\theta)\sqrt{\beta(\theta)}e^{-in\theta}d\theta= \nonumber \\
   &=&\frac{1}{2\pi}\sum_{k=-\infty}^{\infty}u_k
			\int_0^{2\pi}\sqrt{\beta(\theta)}e^{ik\varphi(\theta)-in\theta}d\theta=
         \sum_{k=-\infty}^{\infty} u_kJ_{nk}\,,
\end{eqnarray}
where definition of the matrix $J_{nk}$ is following
\begin{equation}
J_{nk}=\frac{1}{2\pi}\int_0^{2\pi}\sqrt{\beta(\theta)}e^{ik\varphi(\theta)-in\theta}d\theta\,.
\end{equation}
Performing Fourier decomposition of motion equation (\ref{eq:motion}) and using definition of $J_{nk}$ we obtain relation between excitation harmonics $F_m$ ($F=\sum_{m=-\infty}^{\infty} F_m e^{i m\phi}$) and $h_n$ ($h=\sum_{n=-\infty}^{\infty} h_n e^{in\theta}$)
\begin{equation}
F_m=\nu_z\sum_{n=-\infty}^{\infty}J_{nm}^\ast h_n\,.
\end{equation}
Substituting obtained relations into formula (\ref{eq:harmonic-full}) we obtain
\begin{equation}
\displaystyle
\omega_k=-\nu \nu_z\sum_{n,m,l}\frac{D_{kn}n^2J_{nm}J_{lm}^\ast h_l}{\nu_z^2-m^2}\,.
\end{equation}
Calculation of the squared spin harmonic amplitude and averaging over orbits with the same RMS ($\overline{h_l h_s^\ast}=h^2\delta_{ls}$), gives
\begin{equation}
\overline{|\omega_k|^2}=\nu^2\nu_z^2h^2\sum_l |M_{kl}|^2\,,
\end{equation}
where
\begin{equation}
M_{kl}=\sum_n n^2 D_{kn}\sum_m\frac{J_{nm}J_{lm}^\ast}{\nu_z^2-m^2}\,.
\end{equation}
Calculation of the squared distortion harmonic $h^2$ is similar to one performed in previous paragraph, i.e. it is necessary to calculate orbital RMS and perform averaging over orbits. The result is following
\begin{equation}
\overline{\left<z^2\right>}=\nu_z^2h^2\sum_{n,l}
\left|\sum_m\frac{J_{nm}J_{lm}^*}{\nu_z^2-m^2}\right|^2\,
\end{equation}
introducing
\[
Q=\sum_{n,l}\left|\sum_m\frac{J_{nm}J_{lm}^*}{\nu_z^2-m^2}\right|^2\,,
\]
we obtain $h^2=\overline{\left<z^2\right>}/(\nu_z^2Q)$.

\begin{figure}[htb]
\centering
\includegraphics*[width=150mm]{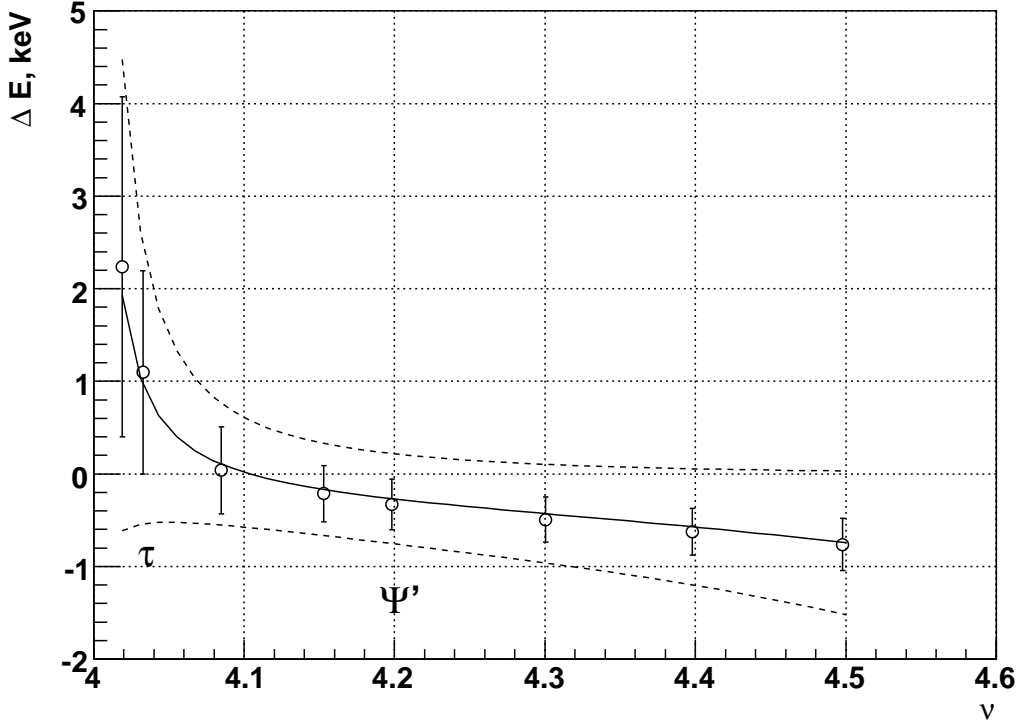}
\caption{Energy shift versus spin tune at 1~mm vertical orbit RMS. The solid line represents estimation by (\ref{eq:result3}), dashed lines represent the uncertainty of the estimate, calculated by (\ref{eq:result3-sigma}), circles with errors are results of the simulation.}
\label{fig:result3}
\end{figure}

Finally, the spin tune shift is
\begin{equation}
\overline{\Delta \nu}=
\frac{\nu^2}{2}\frac{\overline{\left<z^2\right>}}{Q}\sum_{k,l}\frac{|M_{kl}|^2}{\nu-k}\,.
\label{eq:result3}
\end{equation}
In order to calculate the uncertainty of the effect, we have to note that
\[
\overline{h_lh_s^*h_xh_z^*}=
h^4\delta_{ls}\delta_{xz}+h^4\delta_{lz}\delta_{xs}+h^4\delta_{l,-x}\delta_{s,-z}
\]
and $h_l=h_{-l}^*$. Hence, relation for spin harmonics, which defines mean squared spin tune shift is following
\begin{eqnarray}
\overline{|\omega_k|^2\cdot|\omega_t|^2}&=&\nu^4\nu_z^4h^4\times
\left[
\sum_{l,x}|M_{kl}|^2|M_{tx}|^2+\sum_{l,x}M_{kl}M_{tl}^*M_{kx}^*M_{tx}+\right.\nonumber \\
& & \left. +\sum_{l,x}M_{kl}M_{t,-l}^*M_{kx}^*M_{t,-x}^*
\right]\,,
\end{eqnarray}
but the relation for spin harmonics, which defines second power of the mean spin tune shift is
\begin{equation}
\overline{|\omega_k|^2}\cdot\overline{|\omega_t|^2}=\nu^4\nu_z^4h^4
\left[\sum_{l,x}|M_{kl}|^2|M_{tx}|^2\right]\,.
\end{equation}
Performing necessary calculations we obtain the uncertainty of the effect
\begin{equation}
\sigma_{\overline{\Delta\nu}}=\frac{\nu^2}{2}\frac{\overline{\left<z^2\right>}}{Q}
\left[
\displaystyle
\sum_{k,t}\frac{\left|\sum_lM_{kl}M_{tl}^*\right|^2+\left|\sum_lM_{kl}M_{t,-l}\right|^2}
{(\nu-k)(\nu-t)}
\right]^{\frac{1}{2}}\,.
\label{eq:result3-sigma}
\end{equation}
Comparison of the estimation with simulation is presented on Fig.\ref{fig:result3}.

\section{Simulation.}
\begin{figure}[htb]
\centering
\includegraphics*[width=150mm]{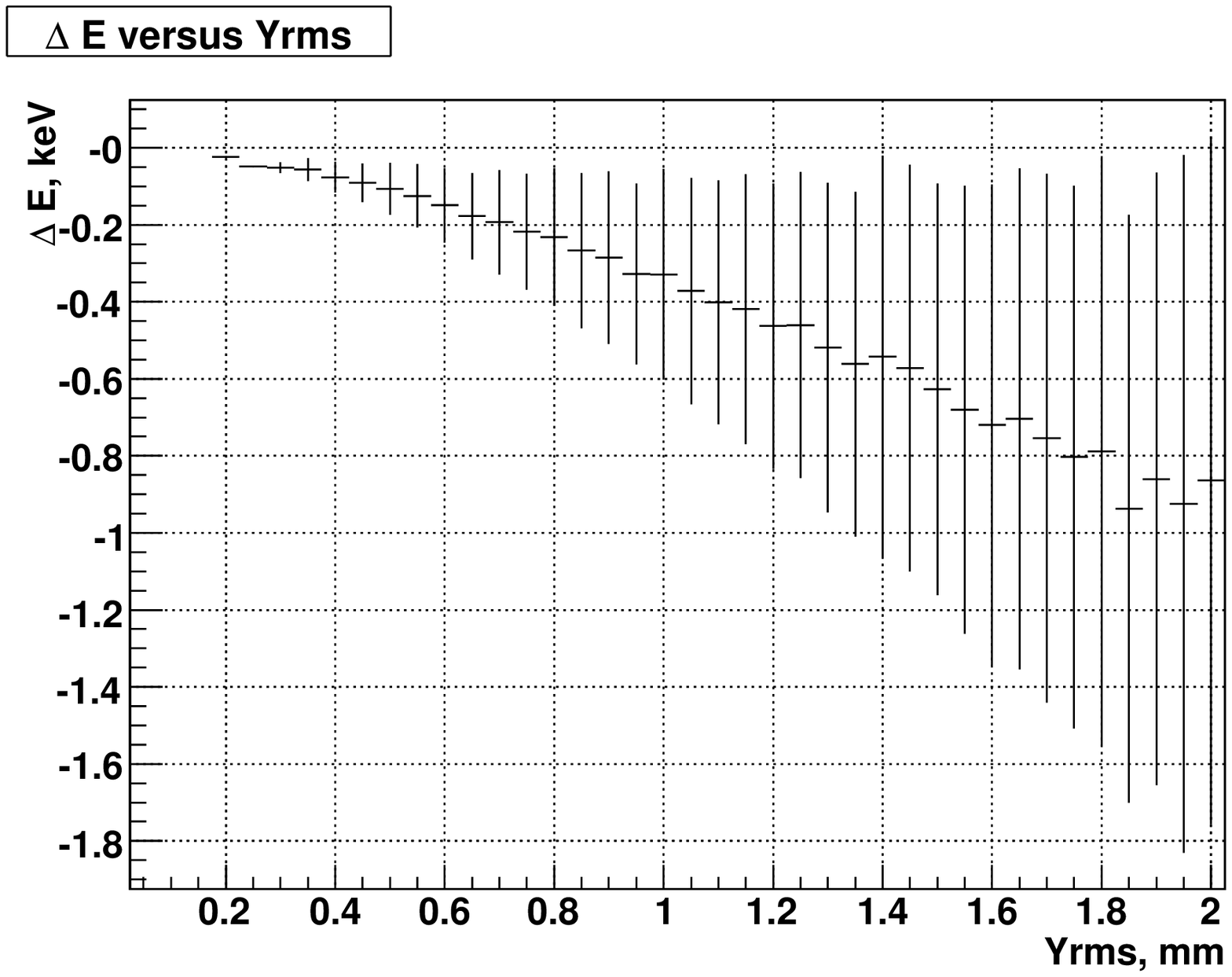}
\caption{Distribution of energy shift $\Delta E$ versus RMS orbit deviation at energy $1850$~MeV. Points are results of simulation.}
\label{fig:dEYrms-psi}
\end{figure}

\begin{figure}[htb]
\centering
\includegraphics*[width=150mm]{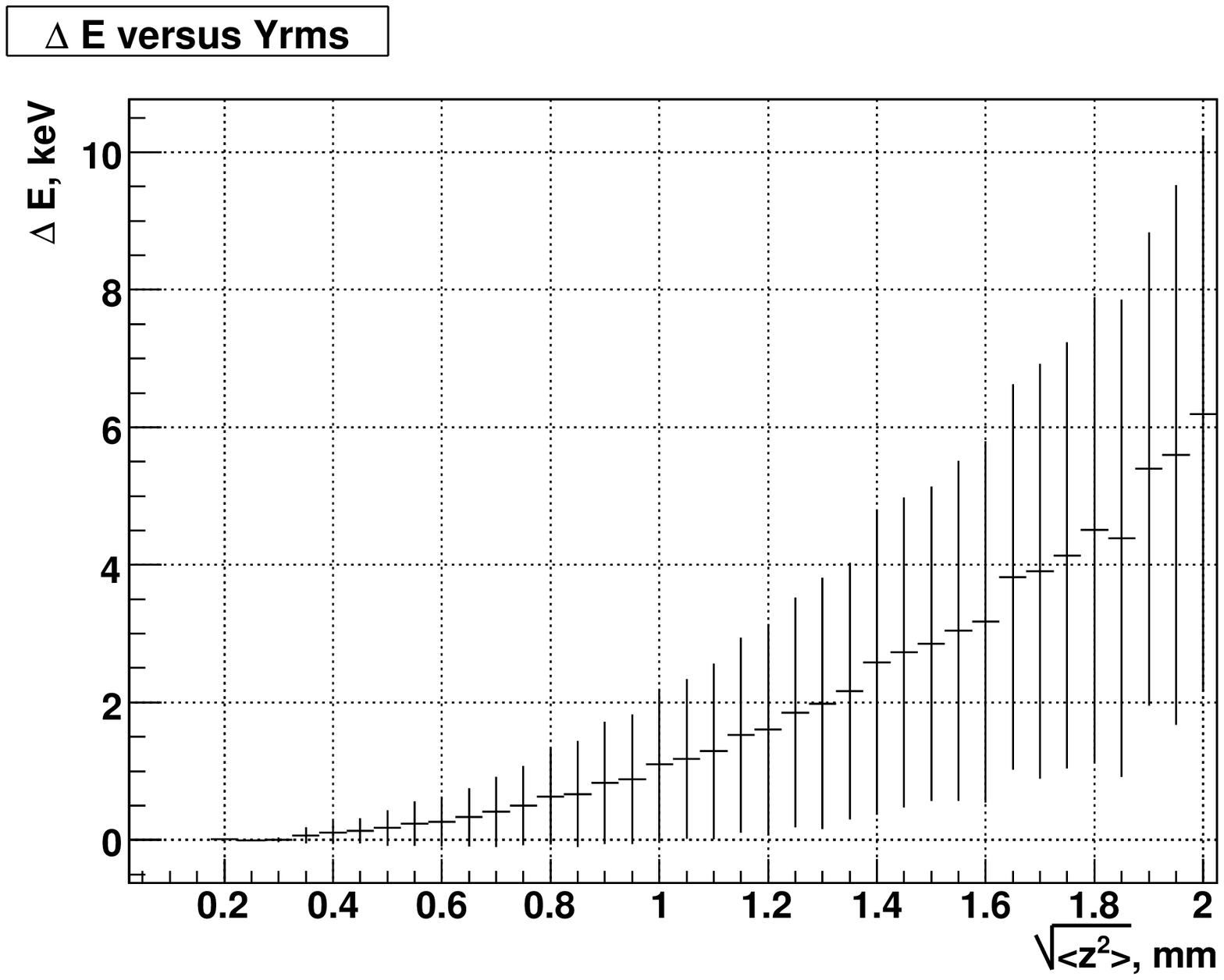}
\caption{Distribution of energy shift $\Delta E$ versus RMS orbit deviation at energy $1777$~MeV. Points are results of simulation.}
\label{fig:dEYrms-tau}
\end{figure}

The Monte-Carlo simulation have been done to understand the possible energy shifts due to vertical closed orbit distortions for VEPP-4M. The sources of the distortions were alignment errors of quadrupoles and sextupoles and random kicks from vertical correctors. For each errors distribution closed orbit has been found and along the closed orbit the spin tune has been calculated using matrix technique. The resulted energy shift $\Delta E=440.64843\cdot(\nu^\prime-\nu)$ is shown on Fig.~\ref{fig:dEYrms-psi}. for $\Psi'$ and on Fig.~\ref{fig:dEYrms-tau}. for $\tau$ lepton.

\subsection{Insufficiency of one harmonic amplitude}
\begin{figure}[htb]
\centering
\includegraphics*[width=150mm]{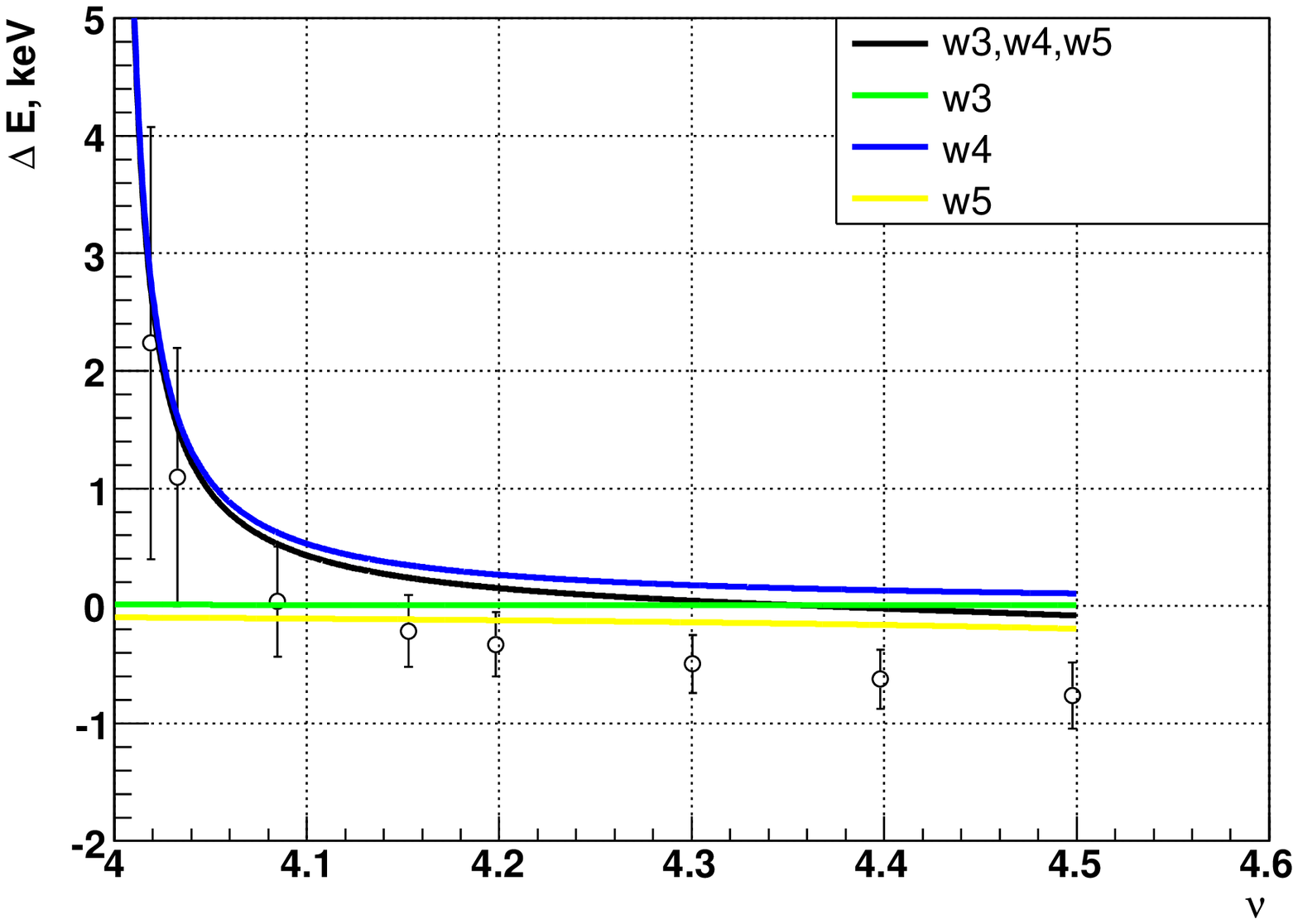}
\caption{Energy shift versus spin tune at 1~mm vertical orbit RMS. Solid lines represent calculations by (\ref{eq:kondratenko}), black---three harmonic consideration, colored---each harmonic separately, circles with errors are results of the simulation.}
\label{fig:har}
\end{figure}

Authors of \cite{Shatunov} proposed usage of single spin harmonic from (\ref{eq:kondratenko}) (measured, for example, by polarization life time) for evaluation the energy bias. To investigate adequacy of such approach three harmonics have been calculated resonant one $\omega_4$ and two harmonics with preceding and subsequent indices $\omega_3$ and $\omega_5$ correspondingly in each simulation run. Then the mean squared harmonic amplitude were found to calculate the spin tune shifts by substituting each harmonic separately and all three together in (\ref{eq:kondratenko}). The comparison of such an approach and simulation is shown on Fig.\ref{fig:har}.
As it could be seen, usage of resonant harmonic approximation is satisfactory in the region not further than 0.1 in units of spin tune. The more distant range requires usage of all other harmonics for correct estimation of the effect.

\section{Conclusion.}
Vertical orbit distortions introduce an energy bias in energy calibration done by resonant depolarization technique. This energy shift could be estimated for general accelerator using assumptions of straight sections absence and constant beta with satisfactory accuracy by formula (\ref{eq:result0-1-sigma}). In case of accelerator with long straight sections, as VEPP-4M, more accurate formula (\ref{eq:result3}) could be used.

The usage of one resonant spin harmonic approximation is adequate only in the immediate region of spin resonance, the area distant on more than 0.1 in units of spin tune from the resonance is not described well by such approach.

In the experiment for $\Psi'$-meson mass measurement the energy shift was $-0.6\pm0.4$~keV with $1.2$~mm of vertical orbit RMS, for $J/\Psi$-meson mass measurement the energy shift was $-0.8\pm0.6$~keV. For the ongoing experiment of $\tau$ lepton mass measurement the effect was estimated to be $1.5\pm1.5$~keV.

\end{document}